\documentclass{llncs}   	
\usepackage{geometry}                		
\usepackage{graphicx}				
\usepackage{amssymb}\usepackage{amsmath}
\begin{document}
\title{Experimental Study of DIGIPASS GO3 and the Security of Authentication }
\titlerunning{Experimental Study of DIGIPASS GO3}
\author{Igor Semaev}\authorrunning{Igor Semaev}
\institute{Department of Informatics, University of Bergen, Norway\\ e-mail: \email{igor@ii.uib.no}}


\maketitle
\begin{abstract}
Based on the analysis of $6$-digit one-time passwords(OTP) generated by DIGIPASS GO3 we were able to reconstruct  the synchronisation system of the token, the OTP generating algorithm and the verification protocol in  details essential for  an attack. The OTPs are more predictable than expected. A forgery attack is described. We argue the attack success probability is $8^{-5}$. That  is much higher than $10^{-6}$ which may be expected if all the digits are independent and uniformly distributed. Under natural assumptions  even in a relatively small bank or company with $10^4$ customers the number of  compromised accounts  during a year may be more than $100$.

\end{abstract}

\section{Introduction}
The remote authentication is commonly multi-way. It is based on a combination of the customer's identifier, his  static password and a dynamic one-time password(OTP) generated by a security token, or  read from a one-time password card, or communicated by a mobile device. 
Static  passwords are at least theoretically may be tapped by a malicious software like trojans with keystroke logging for example, and  used for an attack at some later point. Therefore the authentication security is largely based on the security of the one-time password. Theoretically, the latter is insecure if 
the next one-time password may be  predicted with non-negligible probability given a number of previous one-time passwords. For instance, $6$-digit password may be predicted with probability $10^{-6}$ anyway. If it is possible to predict with a significantly larger probability, then there is an  internal weakness in the generator or in the verification protocol.  The number of attempted wrong log-ins is usually bounded. In practical terms that makes the implementation of the attack more difficult, but does not make that impossible.

 DIGIPASS is a one-time password generator manufactured by VASCO  \cite{Digipass}. In what follows we study $6$-digit combinations produced by DIGIPASS GO3.  According to \cite{Digipass}, the token is  used by lots of customers all over the world, for instance, in Norway, Belgium, Ireland, United Kingdom, Netherlands, Denmark, Saudi Arabia, South Africa etc. It is used  in Norwegian banks as Sparebanken Vest, as described in Sect. \ref{spv}, and DNB NOR. The latter is one of the largest banks in Norway with $2.3$ millions of retail customers and 198000 corporate customers.   The device uses strong crypto algorithms as DES, TDES and AES and supports event and time based authentication. The detailed description of the algorithm was not published.

Based on the experiments with the token, we  show    $6$-digit combinations from DIGIPASS GO3 are dependent and more predictable than expected. The left most digit is used for synchronisation: it indicates a time interval where the token was pressed to generate the  OTP. The left most digit of the current OTP depends on the left most digit of the previous OTP and the time $t$  elapsed since the previous OTP was generated.  The left most digit is predictable with probability almost $1$ if $t$ is close to a multiple of $64$ seconds. Also we found the token may generate the same  combination if $50\leq t<64$ seconds and the probability of that depends on $t$. 
 The last 5 digits of the $6$-digit combination are not uniformly distributed: the digits $0,1,2,3,4,5$ appear with  probability $1/8$, while   $6,7,8,9$ appear with  probability $1/16$. 
   
     We were able to reconstruct the synchronisation system of the token, the OTP generating algorithm and the verification protocol, see Sect. \ref{generate} and Sect. \ref{check}, in  details essential for an attack. Our reconstruction fits well with the experimental data. In particular, it explains  the OTP repetition and the token synchronisation. Though the algorithm implemented in DIGIPASS GO3 is somewhat similar to that published in \cite{RFC6238} and described in Sect. \ref{published}, there is a difference. Due to the synchronisation provided by the left most digit, the verifier  needs only one OTP comparison to authenticate the token.

     A forgery attack is described in Sect. \ref{BasicAttack}. The left most digit does not affect the security of the authentication. Therefore, the  probability of one forged authentication    is $8^{-5}$ instead of 
  $10^{-6}$ as one may expect.

      The attack may be automated and applied against any number of customers which use the token. We study the effect of the attack under natural assumptions that  customer identifiers and static passwords are known to the adversary who may use malicious software as trojans etc to procure them, see for instance \cite{AADK}. Also we assume  the adversary is able to attack after each time the customer uses the token for authentication. Under those assumptions even in a relatively small bank or company with $10^4$ customers the number of  compromised accounts  during a year may be more than $100$.
 
 \section{ Experimental Study of OTPs by DIGIPASS GO3} \label{experiments}
 The experiments were produced by pressing the token and the stopwatch simultaneously   at some time steps.  A very little fluctuation between the real pressing time and that indicated in the tables below   is possible. The data in the tables  was produced by the same DIGIPASS GO3 token. Some of the experiments were repeated for other tokens with similar results.

The customer presses the token to generate an OTP. During the next $40$ seconds this OTP is kept on the  screen and then disappears. If one then presses the token in time between $40$ and $50$ seconds the same OTP reappears. If one presses in  $\geq 50$ seconds a new OTP is generated. The OTP may repeat if the token is pressed in between $50$ and $64$ seconds again as described in the next Sect. \ref{0:50}.
\subsection{ OTPs  at time steps within $[0:50+,1:03+)$}\label{0:50} 
 Tables \ref{50+}-\ref{1:03+} present $6$-digit combinations produced  at  time step  $t$ (format min:sec), where $0:50\leq t<0:51 \,(\mbox{denoted $t=0:50+$}, \mbox{Table}\, \ref{50+} )$, $0:51\leq t<0:52 \,(\mbox{denoted $t=0:51+$}, \mbox{Table}\, \ref{51+} )$ and so on.
  
 For instance, 
29 combinations in Table \ref{50+} were produced  at time step  $t=0:50+$. The first 
 OTP $240445$ indicates starting time  $0:00.00$, the next $302168$ was generated after  $50.35$ sec, the next $498773$ was generated after another $50.41$ sec and so on. Surprisingly, the DIGIPASS happens to repeat the combination. For instance, $642055$ repeats after $0:50+$ seconds again etc. That does not seem  affect the security of the authentication as the server(verifier) does not accept repeated combination as correct.
The sequence of  $6$-digit combinations in the tables   
may be split into intervals, each interval ends with a repeated $6$-digit combination. For $t=0:50+$  the length of the intervals is $4$ or $5$. The right most column in the tables contains a sequence of  differences   modulo 10 between the left most digits in subsequent $6$-digit combinations. That is called a pattern of the sequence. The pattern may be split into intervals accordingly.

  For $t=0:51+$  the length of the intervals is $5$ or $6$. For $t=0:52+$  the length of the intervals is $5$ or $6$ again but the pattern is different. For $t=0:53+$  the length of the intervals is $6$ or $7$ and so on. Finally, 
     for $t=1:03+$
  the length of an interval may be  larger than  $67$.
  
      So   when pressing at  
time step $t$  within $[0:50+,1:03+)$ the first digit of the combination increases by $1$ or is the same and the whole combination repeats. The pattern changes when $t$ changes every second. In particular, the probability that $0$ appears in  the pattern is steadily decreasing while the time step is increasing by  a second. For $t$ around $1:04=64$ seconds the left most digit increases by $1$ with probability very close to $1$.

\subsection{ OTP  at time intervals $t$ within $[1:04+,2:07+)$} 
 When the pressing at  time steps within $[1:04+,2:07+)$ the first digit of the combination increases by $1$ or  $2$ and the pattern changes every second as above.Tables \ref{1:04+}-\ref{2:07+} represent $6$-digit combinations after pressing at time steps $t$, where $t=1:04+,1:21+,1:22+,1:54+,2:07+$.
   The combinations may be split into the intervals. In Tables \ref{1:04+},\ref{1:21+},\ref{1:22+} the interval ends each time when the left most digits increases by $2$ modulo $10$. In Tables \ref{1:54+},\ref{2:07+} the interval ends each time when the left most digits increases by $1$ modulo $10$. 
 For $t=1:21+$ the pattern includes intervals of length $3$ and $4$. For $t=1:22+$ there are    intervals of length $3$ and $4$ again, but the pattern is different. In particular, the  intervals of length $3$ appear more often than the intervals of length $4$. The pattern for $t=1:54+$ is very similar to the pattern for $t=0:50+$ with changing $0,1$ by $1,2$.  
 \subsection{ Pattern in General } \label{pattern}
 Based on the experiments with the time steps $t$ within $[0:50+,1:03+)$ and  within $[1:04+,2:07+)$,  and
 by extrapolating  to any time interval we conclude the following.
 
 \begin{enumerate}
 \item The smallest time interval used in the measurements and  the computations by DIGIPASS and the server is a second.    \item  For 
$(1:04)\times A\leq t< (1:04)\times(A+1)$ the first digit of the combination increases by $A$ or $A+1$ modulo $10$. Each such interval incorporates $64$ seconds and is equivalently represented as 
\begin{equation}\label{basic_interval}
t\in [64\times A ,64\times( A+1)).\end{equation}
 Within (\ref{basic_interval}) the pattern only includes digits $A,A+1$ modulo $10$ and  is changing when $t$ is changing every second.

 \item The patterns are similar for time shifts equal to a multiple of $64$ seconds. In other words, the patterns are  similar  for time steps $t$ and $t+64$ seconds  after substituting $A,A+1$ by  $A+1,A+2$ respectively for any $t$.

 \item Assume an OTP was generated  and   the next OTP  was generated in $t$ seconds. The first digit of the combination may increase by $A$ modulo $10$ only if 
 \begin{equation}\label{basic_interval_1}t\in [64\times(A-1) ,64\times(A+1)).\end{equation}
 
\item The first digit  likely increases by $A-1 $ modulo $10$ in the beginning of the time interval (\ref{basic_interval_1}), by $A+1$ modulo $10$ in the end. In the middle it mostly increases by $A$ modulo $10$. So if $t$ is about  $64\,A$ seconds, the left most digit increases by $A$ modulo $10$ with probability close to $1$.
By symmetry, the probability of getting an increase by $A$ modulo 10 within the interval (\ref{basic_interval_1}) is $1/2$. \end{enumerate}
 Those observations are supported by experiments with random time steps in Table \ref{random} and the time step $10:46$ in Table \ref{10:46+}.

  \subsection{ Synchronising Function}\label{Homomorphism}

Let $(t_0,a_0)$ be some initial values. Let $(t_i,a_i), i=1,2,\dots$ be the sequence of tuples, where $t_i$ is the  time when  an OTP was generated and $a_i$ is its left most digit. 
We claim there is a function $f(t_l,t_k)$ such that for any $l>k\geq 0$ 
\begin{enumerate}
\item  
$f(t_l,t_k)\in\{h_{lk}-1,h_{lk}\},\quad
\mbox{where}\quad
 h_{lk}-1\leq\frac{t_l-t_k}{64}<h_{lk},$
 \item $a_{l}-a_{k}\equiv f(t_{l},t_{k}) \mod 10,$
 \item and
 \begin{equation}\label{sum_f}
f(t_l,t_k)=\sum_{i=k}^{l-1}f(t_{i+1},t_{i}).
\end{equation}
\end{enumerate}
Though the function $f$ may also depend on some other parameters,  that does not affect our conclusions. The value of $f(t_l,t_k)$ represents the number of "time steps" between $t_l$ and $t_k$ and is similar to the time step function used in the well-known time-based OTP algorithm described  in Sect. \ref{published}. However in contrast, $f(t_l,t_k)$ may have two values according to  property 1 of $f$.

 For example, the following Table \ref{random} data was produced by pressing the token for 20 random time intervals $t_{i}-t_{i-1}$ within  1 and 10 minutes or so. The $4$-th column of the table presents integers $h-1,h$ such that   $h-1\leq\frac{t_{i}-t_{i-1}}{64}<h$, where $t_i-t_{i-1}$ was transformed into seconds. The $5$-th column presents the pattern of the sequence of $6$-digit combinations. The $6$-th column gives the value of $f(t_{i},t_{i-1})\in\{h-1,h\}$ according to the definition above.
 \begin{table}[htdp]
 \caption{$6$-digit combinations at random time intervals}
 \begin{center}
\begin{tabular}{|r|c|r|r|c|c|}
\hline
$i$&comb.&$t_{i}-t_{i-1}$&$\frac{t_{i}-t_{i-1}}{64}$&$a_{i}-a_{i-1}\mod 10$&$f(t_{i},t_{i-1})$\\\hline
0&565201&& & &\\
1&057045&6:00& 5,6&5&5\\
2&031320&10:01&9,10&0&10\\
3&317587&3:11&2,3&3&3\\
4&291277&10:05&9,10&9&9\\
5&433132&2:15&2,3&2&2\\
6&911213&5:08&4,5&5&5\\
7&041125&1:17&1,2&1&1\\
8&319430&2:26&2,3&3&3\\
9&253057&10:16&9,10&9&9\\
10&987234&7:31&7,8&7&7\\
11&398564&3:49&3,4&4&4\\
12&070423&7:32&7,8&7&7\\
13&216702&1:43&1,2&2&2\\
14&542368&3:29&3,4&3&3\\
15&914109&4:35&4,5&4&4\\
16&293821&3:28&3,4&3&3\\
17&348346&1:05&1,2&1&1\\
18&913123&5:39&5,6&6&6\\
19&611331&8:10&7,8&7&7\\
20&501416&9:37&9,10&9&9\\\hline
\end{tabular}\quad
\end{center}

\label{random}
\end{table}

 Let us compute $f(t_{11},t_{5})$. As $t_{11}-t_{5}=30:27=1827$ seconds, we have 
$ 28\leq \frac{t_{11}-t_{5}}{64}<29.$ 
By (\ref{sum_f}), we compute
$f(t_{11},t_{5})= 5+1+3+9+7+4= 29$
and see that $a_{11}-a_{5}\equiv f(t_{11},t_{5})  \mod 10.$
Therefore all the properties in the definition of $f$ are satisfied for $f(t_{11},t_{5})$.
Similarly, one computes  $f(t_{l},t_{k})$ for any other $t_{l},t_{k}$ in Table \ref{random} and in any other table of this paper and checks it always satisfies the definition.

\subsection{How does the token generate an OTP} \label{generate}
In this section we reconstruct the way how an OTP is generated by the token. Let $t_0$ be some initial  moment of time and $A_0$  an initial value. 
 Let $t_{i-1}$ be the time of generating the $(i-1)$-th OTP and let $A_{i-1}$ be an auxiliary integer number. Both $t_{i-1}$ and $A_{i-1}$ are kept by the token before the next pair is computed.
 The next OTP $a_i,X_i$ is generated at time $t_{i}, i\geq 1$ by
 \begin{eqnarray}
 A_{i}&=&A_{i-1}+f(t_{i},t_{i-1}),\nonumber\\
a_{i}&\equiv& A_i \mod 10,\nonumber\\ 
X_i&=&E_K(A_i),\nonumber
\end{eqnarray}
where $a_i$ is the OTP left most digit and $X_i$ is a $5$-digit combination, $6$ digits overall. The function $E_K$ is based on an encryption algorithm and depends on a secrete key. By induction and the  property of $f$ in Sect.
\ref{Homomorphism},  
\begin{equation}\label{A}
A_i=A_{i-1}+f(t_{i},t_{i-1})=A_0+f(t_{i-1},t_0)+ f(t_{i},t_{i-1})= A_0+f(t_i,t_0).
\end{equation}
The algorithm above  is similar to one described in Sect. \ref{published}, where the number of time steps is a steadily increasing function: it increases by $1$ after each fixed time interval. In contrast, $f(t_i,t_0)$ may have two values for $t_i$ within the same time interval of $64$ seconds, see Sect. \ref{Homomorphism}. 

 The algorithm fits well with the properties of DIGIPASS GO3 found in Sect. \ref{experiments}. In particular, if $f(t_{i},t_{i-1})=0$, then $A_i=A_{i-1}$ and the whole OTP repeats. The latter is only possible for $t_{i}-t_{i-1}<64$ seconds by the definition of $f$. Also the algorithm explains why the token may be pressed any number of times without authentication and the authentication  still works after, see the next Sect. \ref{check} for details.
\subsection{How does the verifier check the OTP}\label{check}

In what follows we reconstruct the server(verifier) action in authentication. 
When a customer logs in he  introduces his identifier and static passwords into a pop up window on the monitor of his computer. He then presses the DIGIPASS, reads an OTP,  introduces  that into another pop up window  and hits the return key.  When verifying the server is to solve the following three problems.

\begin{enumerate}
\item Handle the delay between generating the OTP and when the server gets it for authentication.
\item  Check if the OTP was produced by the token assigned to that customer.  
 \item Assume  a customer  generates several OTPs   without log in. The server
   should be able to authenticate that customer with the same token  later.

 \end{enumerate}

\subsubsection{Handling a delay.}\label{delay}
 Let  $t$ be the time of generating an OTP and  $t'$ the time it comes to the server for authentication. There should be an acceptable delay time interval $t'-t< T$.  
We found $T$ is  $480$  seconds by the following experiment. An OTP was generated and then introduced to the system with a delay of $t''$ and the server reaction was observed. The results are in Table \ref{Delay}. In this experiment $T$ is $8$ minutes, that is $480$ seconds.  Interestingly,  after the OTP was introduced with a delay of $7:59+$ and accepted, the synchronisation between the DIGIPASS and the server got lost. A new token had to be used.
\begin{table}[htdp]
\caption{Delay handling at the verifier}
\begin{center}
\begin{tabular}{|c|c|c|c|c|c|c|c|c|}
\hline
$t''$&12:40+&10:00+&9:11+&8:00+&7:59+&7:04+&6:02+&5:06+\\\hline
status&reject&reject&reject&reject&accept&accept&accept&accept\\\hline
\end{tabular}
\end{center}
\label{Delay}
\end{table}%
\subsubsection{Defining the value of  $f$ within the delay.}\label{delay}
\noindent Let $t_0$ be an initial time and $A_0$ an initial integer number, both are available for the prover and the verifier. Let  $t$ be the time when the OTP $a,X$ was generated. Let   $t'$ be the time when it comes to the server for verification.
We show the server is able to recover $f(t,t_0)$ if $t'-t< 480$ seconds. Really, 
let $t'$ belong to the interval 
\begin{equation}\label{compute_B}
B-1\leq 
 \frac{t'-t_0}{64}<B
\end{equation} 
for some $B$.
 Therefore if $t'-t<480$ seconds, then
 \begin{equation}\label{compute_B1}
B-9\leq 
 \frac{t-t_0}{64}<B.
\end{equation}
  So $$f(t,t_0)\in \{B-9,\ldots,B-1,B\}$$ 
 by the definition of $f$. As $f(t,t_0)\equiv a-A_0 \mod 10$, the server finds 
 \begin{equation}\label{check_delay}
 f(t,t_0)=A-A_0.
\end{equation}
Simultaneously, the verifier learns an interval of at most $128$ seconds for $t$:
$$(f(t,t_0)-1)\,64\leq t-t_0<(f(t,t_0)+1)\,64.$$
 Moreover it is likely that
$$t\approx f(t,t_0)\,64+t_0$$ 
 seconds by property 5 in Sect.  \ref{pattern}.

\subsubsection{Verification protocol}
In this description we skip the verification of the customer's static passwords.  Let $t$ be the moment of time when the token generates a new OTP $a,X$ and $t'$ be the moment of time when it comes  for verification. The following protocol authenticates the token. 
\begin{enumerate}
\item Compute $B$ by (\ref{compute_B}).
\item Find $f(t,t_0)=A-A_0$ from (\ref{check_delay}) and therefore $A$.

\item  Compute 
  $$X'=E_K(A)$$
and check if $X=X'$. If equality, then the OTP is accepted, otherwise rejected.    
\end{enumerate}
By the properties of $f$ and by (\ref{A}), $A$ only depends on $A_0,t_0$ and $t$. So  the customer may press the token any number of times without log in and that won't affect the next authentication.

\subsection{Last 5 digits  distribution} \label{digits}
  Let $$a\,b\,c\,d\,e\,f$$ be a $6$-digit combinations  produced by DIGIPASS GO3. According to the analysis in previous sections, the first digit $a$ is used for synchronisation. We study the distribution of the other digits $b,c,d,e,f$  taken separately.  
   There are 814 OTPs without repetitions  in all the tables of this article. We count the number of times a decimal digit appears in each of the last five positions in those OTPs. The data is collected in Table \ref{last5}, where the positions are denoted by $b,c,d,e,f$. 
    \begin{table}[htdp]
\caption{The experimental digit distribution  in  the last 5 positions}
\begin{center}
\begin{tabular}{|c|c|c|c|c|c|c|c|c|c|c|}
\hline
 &0&1&2&3&4&5&6&7&8&9\\\hline
b&102&92&94&109&102&105&53&59&48&50\\\hline
c&96&108&96&107&115&80&61&52&45&54\\\hline
d&101&121&109&94&98&103&51&49&47&41\\\hline
e&108&100&108&93&110&97&53&56&45&44\\\hline
f&97&110&110&108&100&83&51&57&54&44\\\hline
\end{tabular}
\end{center}
\label{last5}
\end{table}%
  The distributions are not uniform. The digits $0,1,2,3,4,5$ appear twice more often on the average than $6,7,8,9$. 
   Our explanation is the following. The last $5$ digits are produced from a $20$-bit string of pseudo-random data. One decimal digit per each subsequent $4$-bit string. The latter represents a number from $\{0,1,\ldots,15\}$ and therefore a decimal digit after reduction modulo $10$. As $4$-bit strings are distributed uniformly, the distribution of decimal digits is as $0,1,2,3,4,5$ have probability $1/8$ and $6,7,8,9$ have probability $1/16$.
    That fits well with the experimental  Table \ref{last5} data.

  \section{Basic Attack}  \label{BasicAttack}
We may  assume customer's   static passwords  are  known to the adversary. In what follows a basic algorithm to forge the dynamic password generated by a DIGIPASS is presented and the probability of success is calculated.
Let   $t'$ be the time the latest  OTP generated by the customer came  for authentication, it may be known to the adversary or guessed as well.
\begin{enumerate} 
  \item Take random $a\mod 10$. 
  
  \item Generate $5$ random digits $$b,c,d,e,f\in\{0,1,2,3,4,5\}.$$
  Put $X=b\,c\,d\,e\,f$.
\item     Introduce the forged OTP $a,X$.

\end{enumerate}
To avoid possible security checks at the verifier, one may supply the forged OTP at time $>t'+64$ seconds or so for instance. In order to  authenticate the server constructs a value $A$ such that $A\equiv a \mod 10$, computes $X'=E_K(A)$ and matches $X$ and $X'$.
 
 According to Sect. \ref{digits},  the probability $X=X'$ is $8^{-5}$. 
The probability $p(x)$ of the attack success in case of $x$ independent attempts to log in is  
$$p(x)=1-\left(1-\frac{1}{8^5}\right)^{x}.$$
  
\subsection{One customer is targeted}When the customer finishes his session of authentications, the adversary  communicates  with the server by introducing the customer's identifier and his  static password and tries to    start a new session of authentications with forged OTP according to the algorithm in Section \ref{BasicAttack}. Also there might be a possibility for the adversary to forge  the OTP during the session started by the customer. If not there should be a possibility for him to interrupt the customer's session and then try to start a new session with a forged OTP. Therefore we assume the adversary  is able to start forging the customer's authentication after each finished authentication by the customer.

 For instance, in Norwegian Sparebanken Vest, see Sect. \ref{spv}, each transaction requires an authentication. Let a customer   use the remote authentication once per month to accomplish $10$ authentications, e.g., to start a new session and  to pay $9$ bills. Therefore the number  of forged attempts to authenticate is about $120\,r$ during a year per customer, where $r$ is the number of allowed attempts of authentication. In Sparebanken Vest $r=3$. So the attack success probability applied during a year is then $p(120r)$. If the attacker wants the attack to go unveiled he should wait for a new authentication by the customer to start a series of $r-1$ attempt forged authentication. The 
  success probability is then $p(120(r-1))$ for one year. 
 \begin{table}[htdp]
 \caption{The success probability during a year }
\begin{center}
\begin{tabular}{|c|c|c|c|c|c|c|}
\hline
$r$&1&2&3&4&5&6\\\hline
$p(120\,r)$&0.0036&0.0072&0.0109&0.0145&0.0181&0.0217\\\hline
\end{tabular}

\end{center}
\label{default}
\end{table}%
\subsection{Many customers are targeted}
Let $N$ customers use the remote authentication   to accomplish $10$ authentications per month on the average, that is $120$ authentications  per year. The average number of successful forged authentications is then about 
$Np(120r)$ during one year.  The attack may be automated. If it works against one customer that should work against $N=10^6$ or so  of them. For $r=3$  we have $$Np(120\times 3)\approx N\times 0.010926$$ 
of customer's accounts will get compromised on the average. If $N=10^6$ then the number of compromised accounts is around $10926$ during one year. For $N=10^4$  the number of compromised accounts  during one year is around $109$.

\section{ Authentication in a Norwegian Bank}\label{spv}
There are 3 ways for a customer in Norwegian Sparebanken Vest to authenticate himself and get access to his account for internet banking. They are \begin{enumerate}
\item "BankID-innlogging" requires the social security number, BankID password of the customer's choice and a one-time password from DIGIPASS GO3.  
\item "BankID p\aa\,  mobil" requires the social security number, PIN-code and a mobile phone with a BankID certificate on that.
\item "Alternativ innlogging" requires the social security number, internet banking password of the customer's choice and a one-time password generated by DIGIPASS GO3 or read from a one-time password card.
\end{enumerate}
Two of the three  authentication protocols require a one-time password generator.  

\section{ Published One-Time Password Algorithms}\label{published}
Two One-Time Password(OTP) algorithms  are published in \cite{RFC4226,RFC6238}. The first  HMAC-based One-Time Password (HOTP) algorithm specifies  event-based   
OTP algorithm. The other(TOTP) is en extension of the HOTP algorithm to support the time-based moving factor. We briefly describe the latter. The prover( token) and the verifier(authentication server) use the same time-step value $X$. There must be a unique secret key $K$ for each prover. Therefore,
\begin{equation}\label{eq1}
\textup{TOTP=HOTP($K,T$)}=\mbox{Truncate(HMAC("crypto"},K,T)),\end{equation}
where $T$ is an integer number which represents the number of time steps between the initial counter time $T_0$ and the current Unix time. More specifically,
$$T=\lfloor (\mbox{Current Unix Time} -T_0 )/X\rfloor.$$
 The server should compare OTP not only with the receiving time-stamps but also the past time-stamps that are within a transition delay window, specified by the protocol. 
 The function "Truncate" is specified in   
\cite{RFC4226} for HMAC-SHA-1 as "crypto" in (\ref{eq1}).

\section{Appendices}

 \begin{table}[htdp]
 \caption{$6$-digit combinations at  $0:50+$ }
 \begin{center}
\begin{tabular}{|r|c|c|}
\hline
$i$&comb.&\\\hline
0&240445& \\
1&302168&1\\
2&498773&1\\
3&591327&1\\
4&642055&1\\
5&642055&0\\
6&773743&1\\\hline
\end{tabular}
\quad
\begin{tabular}{|r|c|c|}
\hline

7&818140&1\\
8&943630&1\\
9&943630&0\\

10&091211&1\\
11&147884&1\\
12&220372&1\\
13&388126&1\\
14&388126&0\\\hline
\end{tabular}\quad
\begin{tabular}{|r|c|c|}
\hline
15&475446&1\\
16&534182&1\\
17&621339&1\\
18&747108&1\\
19&747108&0\\
20&829372&1\\
21&904089&1\\
22&012530&1\\\hline
\end{tabular}\quad
\begin{tabular}{|r|c|c|}
\hline

23&141253&1\\
24&141253&0\\
25&251821&1\\
26&370755&1\\
27&455412&1\\
28&455412&0\\\hline
\end{tabular}
\end{center}
\label{50+}
\end{table}

  \begin{table}[htdp]
  \caption{$6$-digit combinations at  $0:51+$}
 \begin{center}
\begin{tabular}{|r|c|c|}
\hline
$i$&comb.&\\\hline
0&613525& \\
1&770281&1\\
2&813348&1\\
3&900173&1\\
4&061082&1\\
5&061082&0\\
6&140542&1\\
7&214421&1\\
8&330592&1\\
9&446002&1\\
10&446002&0\\
11&577502&1\\
12&660042&1\\
13&711359&1\\
14&873243&1\\
15&873243&0\\
16&953483&1\\
17&081487&1\\
18&127404&1\\
19&252525&1\\
20&252525&0\\\hline
\end{tabular}\quad
\begin{tabular}{|r|c|c|}
\hline
21&319740&1\\
22&431246&1\\
23&544489&1\\
24&684950&1\\
25&751093&1\\
26&751093&0\\
27&843031&1\\
28&953144&1\\
29&091433&1\\
30&102531&1\\
31&102531&0\\
32&265373&1\\
33&344032&1\\
34&413491&1\\
35&520503&1\\
36&520503&0\\
37&659320&1\\
38&755504&1\\
39&834672&1\\
40&956211&1\\
41&956211&0\\\hline
\end{tabular}\quad
\begin{tabular}{|r|c|c|}
\hline
42&062056&1\\
43&130310&1\\
44&207562&1\\
45&377484&1\\
46&377484&0\\
47&405446&1\\
48&510028&1\\
49&655673&1\\
50&722642&1\\
51&869254&1\\
52&869254&0\\
53&930220&1\\
54&009236&1\\
55&131738&1\\
56&292465&1\\
57&292465&0\\
58&325694&1\\
59&444324&1\\
60&546511&1\\
61&627135&1\\
62&627135&0\\\hline
\end{tabular}\quad
\begin{tabular}{|r|c|c|}
\hline
63&756821&1\\
64&813067&1\\
65&915018&1\\
66&000513&1\\
67&000513&0\\
68&123010&1\\
69&230168&1\\
70&336542&1\\
71&444814&1\\
72&444814&0\\
73&524691&1\\
74&672561&1\\
75&774225&1\\
76&804713&1\\
77&804713&0\\
78&906800&1\\
79&099042&1\\
80&158124&1\\
81&211271&1\\
82&343838&1\\
83&343838&0\\\hline
\end{tabular}\quad
\end{center}

\label{51+}
\end{table}

\begin{table}[htdp]
\caption{$6$-digit combinations at $0:52+$}
 \begin{center}
\begin{tabular}{|r|c|c|}
\hline
$i$&comb.&\\\hline
0&736910&\\
1&826434&1\\
2&938934&1\\
3&092613&1\\
4&114863&1\\
5&114863&0\\
6&242575&1\\
7&310480&1\\\hline
\end{tabular}\quad
\begin{tabular}{|r|c|c|}
\hline
8&407177&1\\
9&529021&1\\
10&690643&1\\
11&690643&0\\

12&725702&1\\
13&835136&1\\
14&911041&1\\
15&065804&1\\
16&065804&0\\\hline
\end{tabular}\quad
\begin{tabular}{|r|c|c|}
\hline
17&142656&1\\
18&205424&1\\
19&344929&1\\
20&449877&1\\
21&588742&1\\
22&588742&0\\

23&654000&1\\
24&731026&1\\
25&852133&1\\\hline
\end{tabular}\quad
\begin{tabular}{|r|c|c|}
\hline
26&951011&1\\
27&951011&0\\
28&092344&1\\
29&103750&1\\
30&273100&1\\
31&325527&1\\
32&446575&1\\
33&446575&0\\\hline
\end{tabular}\quad
\end{center}

\label{52+}
\end{table}

 \begin{table}[htdp]
 \caption{$6$-digit combinations at  $0:53+$}
 \begin{center}
\begin{tabular}{|r|c|c|}
\hline
$i$&comb.&\\\hline
0&543141&\\
1&632547&1\\
2&750122&1\\
3&820133&1\\
4&971984&1\\
5&023743&1\\
6&023743&0\\
7&123658&1\\
8&213661&1\\\hline
\end{tabular}\quad
\begin{tabular}{|r|c|c|}
\hline

9&304190&1\\
10&402721&1\\
11&584269&1\\
12&584269&0\\

13&693419&1\\
14&781459&1\\
15&873255&1\\
16&909506&1\\
17&034130&1\\
18&034130&0\\\hline
\end{tabular}\quad
\begin{tabular}{|r|c|c|}
\hline

19&146039&1\\
20&212926&1\\
21&324996&1\\
22&412623&1\\
23&518153&1\\
24&518153&0\\

25&696243&1\\
26&703153&1\\
27&807928&1\\\hline
\end{tabular}\quad
\begin{tabular}{|r|c|c|}
\hline

28&913106&1\\
29&024819&1\\
30&024819&0\\
31&103301&1\\
32&261544&1\\
33&367120&1\\
34&488616&1\\
35&534404&1\\
36&683124&1\\
37&683124&0\\\hline
\end{tabular}\quad
\end{center}
\label{53+}
\end{table}

 \begin{table}[htdp]
 \caption{$6$-digit combinations at  $0:54+$}
 \begin{center}
\begin{tabular}{|r|c|c|}
\hline
$i$&comb.&\\\hline
0&460553&\\
1&522574&1\\
2&605910&1\\
3&708033&1\\
4&891210&1\\
5&982543&1\\
6&019508&1\\
7&019508&0\\\hline
\end{tabular}\quad
\begin{tabular}{|r|c|c|}
\hline
8&128426&1\\

9&273175&1\\
10&325114&1\\
11&428472&1\\
12&548721&1\\
13&685321&1\\
14&685321&0\\
15&744745&1\\
16&893417&1\\\hline
\end{tabular}\quad
\begin{tabular}{|r|c|c|}
\hline
17&930741&1\\
18&015819&1\\
19&145630&1\\
20&145630&0\\
21&210016&1\\
22&305020&1\\
23&422742&1\\
24&533817&1\\
25&643375&1\\\hline
\end{tabular}\quad
\begin{tabular}{|r|c|c|}
\hline
26&714284&1\\
27&714284&0\\

28&817215&1\\
29&950261&1\\
30&070143&1\\
31&156623&1\\
32&209504&1\\
33&311863&1\\
34&311863&0\\\hline
\end{tabular}\quad
\end{center}
\label{54+}
\end{table}

 \begin{table}[htdp]
 \caption{$6$-digit combinations at  $0:55+$}
 \begin{center}
\begin{tabular}{|r|c|c|}
\hline
$i$&comb.&\\\hline
0&041004&\\
1&129955&1\\
2&236319&1\\
3&373335&1\\
4&483103&1\\
5&591790&1\\
6&615196&1\\
\hline
\end{tabular}\quad
\begin{tabular}{|r|c|c|}
\hline
7&615196&0\\
8&756362&1\\
9&890229&1\\
10&932420&1\\
11&051043&1\\
12&130620&1\\
13&251594&1\\
14&320845&1\\
\hline
\end{tabular}\quad
\begin{tabular}{|r|c|c|}
\hline
15&320845&0\\
16&424153&1\\
17&589022&1\\
18&640131&1\\
19&784457&1\\
20&820301&1\\
21&900508&1\\
22&900508&0\\\hline

\end{tabular}\quad
\begin{tabular}{|r|c|c|}
\hline
23&063637&1\\
24&172283&1\\
25&215258&1\\
26&324114&1\\
27&440322&1\\
28&511341&1\\
29&649452&1\\
30&649452&0\\\hline
\end{tabular}\quad
\end{center}

\label{55+}
\end{table}

 \begin{table}[htdp]
 \caption{$6$-digit combinations at  $0:56+$}
 \begin{center}
\begin{tabular}{|r|c|c|}
\hline
$i$&comb.&\\\hline
0&232615&\\
1&345751&1\\
2&476543&1\\
3&581103&1\\
4&694218&1\\
5&700474&1\\\hline
\end{tabular}\quad
\begin{tabular}{|r|c|c|}
\hline

6&817790&1\\
7&957312&1\\
8&957312&0\\

9&075022&1\\
10&100474&1\\
11&211555&1\\
12&336028&1\\\hline
\end{tabular}\quad
\begin{tabular}{|r|c|c|}
\hline

13&443090&1\\
14&591452&1\\
15&643205&1\\
16&748041&1\\
17&748041&0\\
18&810307&1\\
19&901680&1\\\hline
\end{tabular}\quad
\begin{tabular}{|r|c|c|}
\hline

20&095142&1\\
21&103354&1\\
22&204011&1\\
23&350105&1\\
24&451132&1\\
25&451132&0\\\hline
\end{tabular}\quad
\end{center}
\label{56+}
\end{table}
 
 \begin{table}[htdp]
 \caption{$6$-digit combinations at  $0:57+$}
 \begin{center}
\begin{tabular}{|r|c|c|}
\hline
$i$&comb.&\\\hline
0&742005&\\
1&801269&1\\
2&906986&1\\
3&026145&1\\
4&152120&1\\
5&273223&1\\
6&369351&1\\
7&454726&1\\
8&572390&1\\
9&572390&0\\
10&606452&1\\
11&741415&1\\
12&830660&1\\
13&956443&1\\\hline

\end{tabular}\quad
\begin{tabular}{|r|c|c|}
\hline

14&041825&1\\
15&119034&1\\
16&201337&1\\
17&305590&1\\
18&452553&1\\
19&452553&0\\

20&546102&1\\
21&617949&1\\
22&794572&1\\
23&806460&1\\
24&904205&1\\
25&042962&1\\
26&172347&1\\
27&234253&1\\
28&234253&0\\\hline

\end{tabular}\quad
\begin{tabular}{|r|c|c|}
\hline

29&302724&1\\
30&437411&1\\
31&510241&1\\
32&682113&1\\
33&752402&1\\
34&812812&1\\
35&914573&1\\
36&069241&1\\
37&133169&1\\
38&133169&0\\
39&287604&1\\
40&344041&1\\
41&409260&1\\
42&536335&1\\
43&673364&1\\\hline

\end{tabular}\quad
\begin{tabular}{|r|c|c|}
\hline

44&754503&1\\
45&859096&1\\
46&915222&1\\
47&044338&1\\
48&044338&0\\
49&165276&1\\
50&262014&1\\
51&304038&1\\
52&409315&1\\
53&504916&1\\
54&642188&1\\
55&756278&1\\
56&807203&1\\
57&807203&0\\\hline
\end{tabular}\quad
\end{center}
\label{57+}
\end{table}
 
 \begin{table}[htdp]
 \caption{$6$-digit combinations at  $0:58+$}
 \begin{center}
\begin{tabular}{|r|c|c|}
\hline
$i$&comb.&\\\hline
0&710424&\\
1&832534&1\\
2&954513&1\\
3&051144&1\\
4&144831&1\\
5&225274&1\\\hline
\end{tabular}\quad
\begin{tabular}{|r|c|c|}
\hline

6&331435&1\\
7&430511&1\\
8&574450&1\\
9&646207&1\\
10&762324&1\\
11&762324&0\\
12&811080&1\\\hline
\end{tabular}\quad
\begin{tabular}{|r|c|c|}
\hline

13&941303&1\\
14&044900&1\\
15&194233&1\\
16&215335&1\\
17&381041&1\\
18&450269&1\\
19&524832&1\\\hline
\end{tabular}\quad
\begin{tabular}{|r|c|c|}
\hline

20&611601&1\\
21&714464&1\\
22&876142&1\\
23&876142&0\\\hline
\end{tabular}\quad
\end{center}
\label{58+}
\end{table}
 
 \begin{table}[htdp]
 \caption{$6$-digit combinations at  $0:59+$}
 \begin{center}
\begin{tabular}{|r|c|c|}
\hline
$i$&comb.&\\\hline
0&132935&\\
1&279542&1\\
2&344200&1\\
3&413388&1\\
4&578023&1\\
5&085158&1\\
6&731544&1\\
7&815031&1\\
8&945111&1\\
9&064830&1\\\hline
\end{tabular}\quad
\begin{tabular}{|r|c|c|}
\hline

10&171118&1\\
11&231314&1\\
12&343154&1\\
13&421393&1\\
14&421393&0\\

15&553101&1\\
16&635170&1\\
17&759238&1\\
18&815107&1\\
19&965207&1\\
20&000298&1\\\hline
\end{tabular}\quad
\begin{tabular}{|r|c|c|}
\hline

21&109515&1\\
22&266521&1\\
23&344521&1\\
24&456243&1\\
25&557915&1\\
26&612537&1\\
27&786251&1\\
28&786251&0\\

29&836131&1\\
30&957731&1\\
31&051304&1\\\hline
\end{tabular}\quad
\begin{tabular}{|r|c|c|}
\hline

32&108404&1\\
33&275774&1\\
34&353233&1\\
35&404253&1\\
36&578471&1\\
37&646855&1\\
38&712803&1\\
39&849321&1\\
40&921401&1\\
41&061879&1\\
42&061879&0\\\hline

\end{tabular}\quad
\end{center}
\label{59+}
\end{table}

   \begin{table}[htdp]
   \caption{$6$-digit combinations at  $1:00+$}
 \begin{center}
\begin{tabular}{|r|c|c|}
\hline
$i$&comb.&\\\hline
0&724454&\\
1&861432&1\\
2&907425&1\\
3&033217&1\\
4&126058&1\\
5&235589&1\\\hline
\end{tabular}\quad
\begin{tabular}{|r|c|c|}
\hline

6&323641&1\\
7&454143&1\\
8&454143&1\\
9&580265&1\\
10&648310&1\\
11&752930&1\\
12&864301&1\\\hline
\end{tabular}\quad
\begin{tabular}{|r|c|c|}
\hline

13&940556&1\\
14&335577&0\\
15&162684&1\\
16&285611&1\\
17&391576&1\\
18&422613&1\\
19&558204&1\\\hline
\end{tabular}\quad
\begin{tabular}{|r|c|c|}
\hline

20&693002&1\\
21&711011&1\\
22&843759&1\\
23&959331&1\\
24&019107&1\\
25&192847&1\\
26&192847&1\\\hline
\end{tabular}\quad
\end{center}
\label{1:00+}
\end{table}

 \begin{table}[htdp]
 \caption{$6$-digit combinations at  $1:01+$}
 \begin{center}
\begin{tabular}{|r|c|c|}
\hline
$i$&comb.&\\\hline
0&803133&\\
1&922672&1\\
2&062016&1\\
3&141753&1\\
4&276554&1\\
5&315994&1\\
6&401737&1\\
7&590233&1\\
8&634440&1\\
9&746417&1\\
10&844471&1\\
11&927342&1\\
12&023403&1\\\hline
\end{tabular}\quad
\begin{tabular}{|r|c|c|}
\hline

13&133123&1\\
14&243252&1\\
15&327112&1\\
16&464543&1\\
17&573035&1\\
18&665709&1\\
19&731074&1\\
20&895420&1\\
21&973368&1\\
22&065620&1\\
23&181246&1\\
24&232721&1\\
25&304903&1\\
26&440132&1\\\hline
\end{tabular}\quad
\begin{tabular}{|r|c|c|}
\hline

27&440132&0\\
28&509292&1\\
29&680841&1\\
30&750414&1\\
31&860648&1\\
32&986760&1\\
33&034146&1\\
34&184532&1\\
35&221552&1\\
36&348100&1\\
37&405404&1\\
38&527308&1\\
39&641451&1\\
40&707232&1\\\hline
\end{tabular}\quad
\begin{tabular}{|r|c|c|}
\hline

41&841782&1\\
42&954933&1\\
43&028616&1\\
44&157851&1\\
45&253403&1\\
46&316240&1\\
47&453212&1\\
48&533493&1\\
49&610200&1\\
50&754655&1\\
51&840223&1\\
52&944388&1\\
53&944388&0\\
\hline
\end{tabular}\quad
\end{center}

\label{1:01+}
\end{table}

 \begin{table}[htdp]
 \caption{$6$-digit combinations at  $1:02+$}
 \begin{center}
\begin{tabular}{|r|c|c|}
\hline
$i$&comb.&\\\hline
0&049789&\\
1&125454&1\\
2&249724&1\\
3&336253&1\\
4&434262&1\\
5&534011&1\\
6&634963&1\\
7&707424&1\\
8&864346&1\\
9&904542&1\\
10&055374&1\\
11&109826&1\\
12&250174&1\\
13&393648&1\\
14&472334&1\\
15&522314&1\\\hline
\end{tabular}\quad
\begin{tabular}{|r|c|c|}
\hline
16&634277&1\\

17&752091&1\\
18&810267&1\\
19&918023&1\\
20&045143&1\\
21&110766&1\\
22&250180&1\\
23&390264&1\\
24&407684&1\\
25&407684&0\\
26&517745&1\\
27&621362&1\\
28&741133&1\\
29&825148&1\\
30&921056&1\\
31&021283&1\\
32&132300&1\\
\hline
\end{tabular}\quad
\begin{tabular}{|r|c|c|}
\hline
33&268803&1\\
34&374990&1\\
35&414499&1\\
36&541519&1\\
37&687152&1\\
38&705422&1\\
39&879564&1\\
40&988056&1\\
41&038402&1\\
42&184004&1\\
43&203522&1\\

44&345529&1\\
45&465269&1\\
46&502087&1\\
47&603136&1\\
48&724034&1\\

49&821852&1\\\hline
\end{tabular}\quad
\begin{tabular}{|r|c|c|}
\hline
50&936521&1\\
51&017643&1\\
52&107207&1\\
53&253850&1\\
54&308923&1\\
55&442758&1\\
56&535311&1\\
57&694302&1\\
58&704092&1\\
59&808352&1\\
60&956510&1\\
61&053083&1\\
62&154862&1\\
63&223165&1\\
64&372593&1\\
65&372593&0\\
\hline
\end{tabular}\quad
\end{center}

\label{1:02+}
\end{table}

 \begin{table}[htdp]
 \caption{$6$-digit combinations at  $1:03+$}
 \begin{center}
\begin{tabular}{|r|c|c|}
\hline
$i$&comb.&\\\hline
0&197308&\\
1&229653&1\\
2&314183&1\\
3&314183&0\\
4&472242&1\\
5&548311&1\\
6&653232&1\\
7&751238&1\\
8&853136&1\\
9&994925&1\\
10&003939&1\\
11&136055&1\\
12&242150&1\\
13&377020&1\\
14&433284&1\\
15&589263&1\\
16&697642&1\\\hline
\end{tabular}\quad
\begin{tabular}{|r|c|c|}
\hline
17&736565&1\\
18&877010&1\\
19&932921&1\\
20&001830&1\\
21&117801&1\\
22&233531&1\\
23&390642&1\\
24&414560&1\\
25&519185&1\\
26&623444&1\\
27&726137&1\\
28&815144&1\\
29&913133&1\\
30&050445&1\\
31&154051&1\\
32&205540&1\\
33&371497&1\\
34&428420&1\\\hline
\end{tabular}\quad
\begin{tabular}{|r|c|c|}
\hline
35&562882&1\\

36&696411&1\\
37&793512&1\\
38&856523&1\\
39&954494&1\\
40&001432&1\\
41&159710&1\\
42&253131&1\\
43&343330&1\\
44&453233&1\\
45&564870&1\\
46&604627&1\\

47&775411&1\\
48&842291&1\\
49&967451&1\\
50&005706&1\\
51&102313&1\\
52&240554&1\\\hline
\end{tabular}\quad
\begin{tabular}{|r|c|c|}
\hline
53&354324&1\\
54&450806&1\\
55&551977&1\\
56&682610&1\\
57&710400&1\\
58&830638&1\\
59&963248&1\\
60&059241&1\\
61&110925&1\\
62&217217&1\\
63&344198&1\\
64&407524&1\\
65&552941&1\\
66&631306&1\\
67&764802&1\\
68&880970&1\\
69&921905&1\\
70&063039&1\\
\hline
\end{tabular}\quad
\end{center}

\label{1:03+}
\end{table}

 \begin{table}[htdp]
 \caption{$6$-digit combinations at  $1:04+$}
 \begin{center}
\begin{tabular}{|r|c|c|}
\hline
$i$&comb.&\\\hline
0&269612&\\
1&352122&1\\
2&430378&1\\
3&501426&1\\
4&624726&1\\
5&770000&1\\
6&858070&1\\
7&954795&1\\
8&054725&1\\
9&101035&1\\
10&215597&1\\
11&324103&1\\
12&431417&1\\
13&566145&1\\
14&624507&1\\
15&744115&1\\\hline
\end{tabular}\quad
\begin{tabular}{|r|c|c|}
\hline
16&803105&1\\
17&964970&1\\
18&085128&1\\
19&149115&1\\
20&202403&1\\
21&393555&1\\

22&434318&1\\
23&544247&1\\
24&900020&1\\
25&075758&1\\
26&131647&1\\
27&236059&1\\
28&361743&1\\
29&487607&1\\
30&550510&1\\
31&640136&1\\
32&788015&1\\\hline
\end{tabular}\quad
\begin{tabular}{|r|c|c|}
\hline

33&834055&1\\
34&952044&1\\
35&006042&1\\
36&114194&1\\
37&200429&1\\
38&332250&1\\
39&434921&1\\
40&541209&1\\
41&628292&1\\
42&721117&1\\
43&845051&1\\

44&922050&1\\
45&024254&1\\
46&189460&1\\
47&203522&1\\
48&333231&1\\
49&432931&1\\\hline
\end{tabular}\quad
\begin{tabular}{|r|c|c|}
\hline
50&525517&1\\
51&632156&1\\
52&701365&1\\
53&823442&1\\
54&967919&1\\
55&057862&1\\
56&117521&1\\
57&234149&1\\
58&315074&1\\
59&452005&1\\
60&530501&1\\
61&633500&1\\
62&720237&1\\
63&898292&1\\
64&906445&1\\
65&155602&2\\
\hline
\end{tabular}\quad
\end{center}
\label{1:04+}
\end{table}

\begin{table}[htdp]
\caption{$6$-digit combinations at  $1:21+$}
 \begin{center}
\begin{tabular}{|r|c|c|}
\hline
$i$&comb.&\\\hline
0&729021&\\
1&862954&1\\
2&984329&1\\
3&027332&1\\
4&283083&2\\
5&354734&1\\
6&430327&1\\
7&620593&2\\
8&732308&1\\
9&830515&1\\\hline
\end{tabular}\quad
\begin{tabular}{|r|c|c|}
\hline
10&988222&1\\
11&124510&2\\
12&201595&1\\
13&320271&1\\
14&444368&1\\
15&648544&2\\
16&730542&1\\
17&888677&1\\
18&011941&2\\
19&135073&1\\
20&272046&1\\\hline
\end{tabular}\quad
\begin{tabular}{|r|c|c|}
\hline
21&334242&1\\
22&530457&2\\
23&672523&1\\
24&739122&1\\
25&904289&2\\
26&066584&1\\
27&126007&1\\
28&252851&1\\
29&426375&2\\
30&522133&1\\
31&602702&1\\\hline
\end{tabular}\quad
\begin{tabular}{|r|c|c|}
\hline
32&759044&1\\
33&993272&2\\
34&041015&1\\
35&188531&1\\
36&338403&2\\
37&419244&1\\
38&520143&1\\
39&623183&1\\
40&834127&2\\\hline
\end{tabular}\quad
\end{center}

\label{1:21+}
\end{table}

\begin{table}[htdp]
\caption{$6$-digit combinations at  $1:22+$}
 \begin{center}
\begin{tabular}{|r|c|c|}
\hline
$i$&comb.&\\\hline
0&373763&\\
1&406375&1\\
2&533966&1\\
3&704802&2\\
4&875121&1\\
5&927450&1\\
6&032654&1\\
7&299538&2\\
8&325157&1\\
9&441409&1\\
10&650483&2\\
11&717743&1\\
12&883525&1\\\hline
\end{tabular}\quad
\begin{tabular}{|r|c|c|}
\hline
13&965143&1\\
14&112015&2\\
15&237007&1\\
16&331278&1\\
17&515093&2\\

18&640216&1\\
19&751552&1\\
20&858301&1\\
21&015311&2\\
22&126288&1\\
23&200451&1\\
24&421201&2\\
25&593095&1\\
26&635180&1\\\hline
\end{tabular}\quad
\begin{tabular}{|r|c|c|}
\hline
27&841140&2\\
28&992450&1\\
29&081502&1\\
30&102493&1\\
31&312450&2\\
32&439123&1\\
33&516114&1\\
34&772509&2\\
35&864425&1\\
36&963344&1\\
37&073140&1\\
38&209594&2\\
39&336225&1\\
40&443141&1\\\hline
\end{tabular}\quad
\begin{tabular}{|r|c|c|}
\hline
41&675541&2\\
42&710104&1\\
43&863144&1\\
44&953202&1\\
45&155572&2\\
46&243962&1\\
47&391080&1\\
48&534333&2\\
49&642176&1\\
50&772069&1\\
51&857507&1\\
52&008457&2\\\hline

\end{tabular}\quad
\end{center}

\label{1:22+}
\end{table}

  \begin{table}[htdp]
  \caption{$6$-digit combinations at  $1:54+$}
 \begin{center}
\begin{tabular}{|r|c|c|}
\hline
$i$&comb.&\\\hline
0&834101& \\
1&090148&2\\
2&292408&2\\
3&428030&2\\
4&531451&1\\
5&766274&2\\
6&941733&2\\
7&135251&2\\
8&375411&2\\
9&430301&1\\\hline
\end{tabular}\quad
\begin{tabular}{|r|c|c|}
\hline
10&655822&2\\
11&810212&2\\
12&050308&2\\
13&215335&2\\
14&310162&1\\
15&595105&2\\
16&749532&2\\
17&935564&2\\
18&028050&1\\
19&203259&2\\
20&488972&2\\\hline
\end{tabular}\quad
\begin{tabular}{|r|c|c|}
\hline
21&674041&2\\
22&802045&2\\
23&985324&1\\
24&122040&2\\
25&355406&2\\
26&523265&2\\
27&751112&2\\
28&860390&1\\
29&009811&2\\
30&203422&2\\
31&430329&2\\\hline
\end{tabular}\quad
\begin{tabular}{|r|c|c|}
\hline
32&504147&1\\
33&739322&2\\
34&940505&2\\
35&126602&2\\
36&324350&2\\
37&487094&1\\
38&639081&2\\
39&851054&2\\
40&053215&2\\
41&250881&2\\
42&362597&1\\\hline
\end{tabular}\quad
\end{center}
\label{1:54+}
\end{table}

\clearpage

\begin{table}[htdp]
\caption{$6$-digit combinations at  $2:07+$}
 \begin{center}
\begin{tabular}{|r|c|c|}
\hline
$i$&comb.&\\\hline
0&273335& \\
1&483158&2\\
2&641051&2\\
3&803836&2\\\hline
\end{tabular}\quad
\begin{tabular}{|r|c|c|}
\hline
4&000508&2\\
5&296752&2\\
6&431516&2\\
7&643494&2\\
8&833238&2\\\hline
\end{tabular}\quad
\begin{tabular}{|r|c|c|}
\hline
9&000822&2\\
10&114160&1\\
11&379050&2\\
12&571410&2\\
13&781454&2\\\hline
\end{tabular}\quad
\begin{tabular}{|r|c|c|}
\hline
14&900609&2\\
15&154812&2\\
16&303585&2\\
17&529381&2\\\hline
\end{tabular}\quad
\end{center}
\label{2:07+}
\end{table}

\begin{table}[htdp]
\caption{$6$-digit combinations at  $10:46+$}
 \begin{center}
\begin{tabular}{|r|c|c|}
\hline
$i$&comb.&\\\hline
0&452170&0\\
1&421660&0\\
2&462345&0\\
3&431757&0\\\hline
\end{tabular}\quad
\begin{tabular}{|r|c|c|}
\hline
4&461315&0\\
5&432355&0\\
6&420151&0\\
7&431628&0\\
8&422360&0\\\hline
\end{tabular}\quad
\begin{tabular}{|r|c|c|}
\hline
9&576113&1\\
10&559829&0\\
11&513224&0\\
12&540014&0\\
13&523659&0\\\hline
\end{tabular}\quad
\begin{tabular}{|r|c|c|}
\hline
14&572212&0\\
15&598353&0\\
16&596342&0\\
17&531571&0\\\hline
\end{tabular}\quad
\end{center}

\label{10:46+}
\end{table}

\end{document}